# Haptic Simulator for Liver Diagnostics through Palpation

Felix G. Hamza-Lup[a,1], Crenguta M. Bogdan[b], Adrian Seitan[b]
[a]*Computer Science and Information Technology, Armstrong Atlantic State University*
[b]*Mathematics and Informatics Faculty, Ovidius University, Constanta*

**Abstract.** Mechanical properties of biological tissue for both histological and pathological considerations are often required in disease diagnostics. Such properties can be simulated and explored with haptic technology. Development of cost effective haptic-based simulators and their introduction in the minimally invasive surgery learning cycle is still in its infancy. Receiving pre-training in a core set of surgical skills can reduce skill acquisition time and risks. We present the development of a visuo-haptic simulator module designed to train internal organs disease diagnostics through palpation. The module is part of a set of tools designed to train and improve basic surgical skills for minimally invasive surgery.

**Introduction**

In the area of medical diagnostics and minimally invasive surgery there is a strong need to determine mechanical properties of biological tissue for both histological and pathological considerations. One of the established diagnosis procedures is the palpation of body organs and tissue. In laparoscopic surgery internal tissue palpation is an important pre-operatory activity [1] and an important step in disease diagnostics. Moreover safe practice requires surgeons to respond correctly to both visual and haptic cues.

The focus of this work is the development and assessment of an affordable visuo-haptic simulator built with off-the-shelf components and designed to improve practice-based education in minimally invasive surgery. We present the diagnostics through palpation module in conjunction with a 3D deformable liver model and a few associated disease cases.

In the last decade several frameworks, APIs and toolkits were proposed and implemented to facilitate haptic interface development [2-4]. The simulator was developed with an open source API for haptics (H3D) and using an open standard for the visual 3D component (X3D).

---

[1] Felix G. Hamza-Lup. Computer Science and Information Technology, Armstrong Atlantic State University. E-mail: Felix.Hamza-Lup@armstrong.edu

## 1. Liver Diagnostics through Palpation

The liver is the largest organ in the human body. During lifetime, liver size increases with increasing age, averaging five centimeters span at five years and attaining adult size by age fifteen. The size depends on several factors: age, sex, body size and shape, as well as the particular examination technique utilized (e.g., palpation versus percussion versus radiographic). A liver span two to three cm larger or smaller than these values is considered abnormal [5].

The normal liver is smooth, with no irregularities. In laparoscopic surgery internal tissue palpation is an important pre-operatory activity [6]. Liver palpation can reveal multiple issues: presence of emphysema with an associated depressed diaphragm, fatty infiltration (enlarged with rounded edge), active hepatitis (enlarged and tender), cirrhosis (enlarged with nodular irregularity), hepatic neoplasm (enlarged with rock-hard or nodular consistency).

## 2. Haptic Simulation Module for Liver Palpation

The simulator implementation uses a deformable mesh model. Two important parameters when implementing a deformable 3D object are the object's surface properties and its stiffness. The first one defines the visual and haptic properties at touch while the second one is part of the deformation algorithm that provides visual as well as force feedback during object palpation.

The liver 3D model uses the X3D [7] triangle set. The initial 40k polygonal model is reduced to a 3k polygonal model (Figure 1) in order to obtain smooth real-time visuo-haptic deformation behavior using the H3D [2] platform while preserving the resemblance with a real liver. Several models were necessary to simulate disease conditions. We started with a model describing the elastic properties of a healthy liver. The deformation algorithm for the healthy liver is based on a spring-damper model, meaning that the haptic device will render a force directly proportional with the penetration distance into the deformable object. Adjustments to this model are made based on the surgeons' feedback and based on the measurements done on a real liver (i.e. we used a pork liver sample) with a pen shaped force meter. In addition to the deformation model for a healthy liver, the simulator also provides several deformation models corresponding to various disease scenarios and states e.g. liver with cists, cirrhosis etc. The malignant tissue is simulated using 3D polygonal shapes with a different stiffness value than the liver tissue.

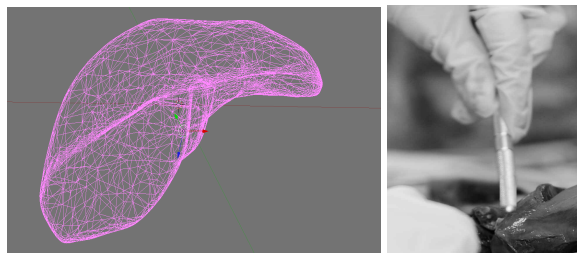

**Figure 1.** Wireframe 3D deformable liver model (left), force-meter on the liver sample (right)

The training scenario starts with a basic haptic task to get the student accustomed with the haptic interface. An interactive assessment of the student's haptic manipulation skills prevents him from moving to the next step before his motor skills are adjusted based on the haptic and visual feedback. Next, the following scenarios are presented to the student in a random order: healthy liver, liver with cirrhosis, liver with tumors – malign and/or cysts, hepatic liver and enlarged liver. Figure 2 illustrates the healthy liver scenario. At the end of each scenario a timed questionnaire is used to assess the student's capability to correctly identify the various healthy/diseased liver scenarios. In each scenario the student has access to a set of tools that s/he can use for palpation (e.g. Babcock grasper, Maryland grasping forceps). The main goal is to diagnose and use the information gathered to decide on the set of actions/procedures required for the next step of the operation.

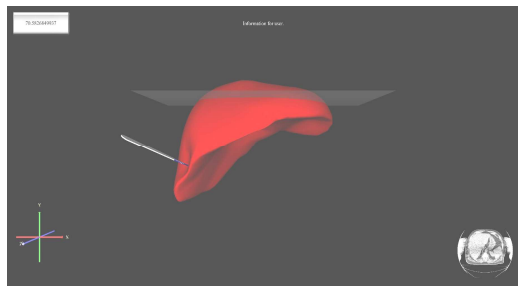

**Figure 2.** Haptic liver palpation simulator with CT scan.

The palpation can be executed through gently pushing on certain parts of the liver or by pressing and following the liver surface. The force on the liver surface is constantly recorded and the peaks are reported in the assessment file at the end of the simulation. If the palpation force increases close to a certain threshold the student is warned through a pop-up message. If the force exceeds the threshold, the simulation stops and the student fails the current scenario since the liver tissue may have been damaged. The associated set of CT scans is presented in one corner of the screen and the student can navigate through it in parallel with the haptic analysis. The current section plan of the CT scan set is projected in 3D as a mesh overlapping the liver model at the correct image location as illustrated in Figure 2.

The force-feedback hardware interface is implemented through a Phantom Omni device from Senseable™ Technologies (Figure 3). The 3D visualization system relies on a pair of NVidia 3D shutter glasses synchronized with an NVidia GeForce GPU.

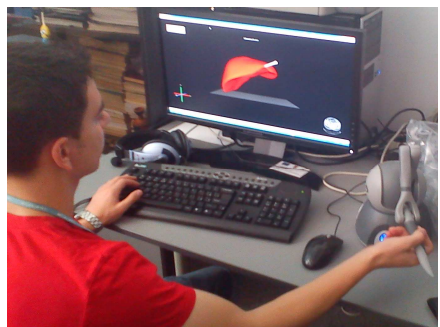

**Figure 3.** Student experimenting with the haptic liver palpation simulator.

## 3. Simulator Assessment

Cost minimization is high priority on hospital agendas; however training and assessment under the apprenticeship model is expensive as it increases procedure learning time. Currently most of the surgery residents' skill evaluation is performed by expert surgeons. This makes the evaluation process costly and sometimes subjective. Using a well designed visuo-haptic simulation system which supports skill assessment reduces this subjectivity issue and the probability of human error.

"Methods for objective assessment of technical proficiency of surgeons are explored in [8] and an important taxonomy of metrics for the evaluation of surgical abilities and skills is presented by Satava in [9]." This taxonomy is based on two main concepts: validity and reliability. Each test is designed for a specific objective. The first concept, *validity* of a test, refers to accepting a test if it is in compliance with five validity measures. The second concept, *reliability* of a test, refers to the consistency of the results as the test is performed multiple times by the same person or by different persons. Five validity measures are defined: face, content, construct, concurrent and predictive. These validity metrics endorse the test's fulfillment of the objective. Each metric determines the objective fulfillment from a different perspective:

- *Face validity* is determined by the appearance of the interface of the simulated task addressed by the test.
- *Content validity* is determined by the expert surgeons based on the detailed examination of the test content.
- *Construct validity* is determined by the capability of the test to differentiate among performance levels.
- *Concurrent validity* is determined by the capability of the test to return equivalent results with other similar tests.
- *Predictive validity* is determined by the predictive capability of the test. The evaluated resident will have the same performance

Two complementary metrics are defined for test reliability. *Inter-rater reliability* - when the test is performed by two independent evaluators, their results are sufficiently close (if not similar) and *Test-retest reliability* - repeating the test at different times and dates should return comparable results for multiple evaluators. Formal validation studies that focus on the use of haptics in medical simulation are scarce, with most simulations only assessing the simulation's face validity as the other metrics are quite time consuming to evaluate and their effect is difficult to objectively prove.

We have employed the *validity* metric for the liver palpation simulator with the following meanings. *Face validity*: is determined by the visuo-haptic characteristics of the interface (i.e., how the simulated 3D liver looks and feels in comparison with the real liver). *Content validity*: if the test measures a certain skill, in this case correct diagnostics through liver palpation. *Construct validity*: the test results should be able to allow differentiation between an expert and a novice surgeon. *Concurrent validity*: the capability of a test to return equivalent results with other similar test for the same skill. *Predictive validity*: certainty that, after passing the test, the resident will have similar performance in a real environment.

During the development phase we focused on face validity and content validity. Face validity was measured objectively as well as subjectively. These preliminary measures will allow us to fine-tune both the application as well as the assessment strategy. For the objective assessment we employed a force meter to measure forces applied on the liver tissue and compare them with the forces generated by the simulator. The visual component requires color calibration and it is usually part of the subjective assessment. The content validity is measured objectively through the questionnaires presented at the end of each scenario as described in Section 2. For each validity metric we have used a score that ranges from 1 to 10, 10 being the highest score. The scores obtained in the preliminary measurement based on the analysis of one expert surgeon and two resident students are 8 for the face validity and a 9 for content validity.

At the time of this writing the assessment of the simulator is still in progress. We have several resident students at the Constanta Regional Hospital, Internal Surgery Division, currently using the simulator and providing feedback.

## 4. Conclusions

Palpation in minimally invasive surgery is where a practitioner presses upon an area of interest with a tool to locate landmarks and feel for the presence or absence of anatomic and/or physiological features or abnormalities. Training of such an intensive tactile skill can be achieved through haptic enabled simulators. We presented a visuo-haptic simulation module and provided a preliminary assessment for face and content validity. A complete evaluation of transfer of skills where one control group does not use a simulator and another uses the simulator with haptic feedback has yet to be carried out.

### Acknowledgments

This study was supported under the ANCS grant "HapticMed – Using haptic interfaces in medical applications", no. 128/02.06.2010, ID/SMIS 567/12271.